\documentstyle[12pt]{article}   
\begin{document}

\title{{\small\centerline{November 1994 \hfill IIA-NAPP-94-13}}
\medskip
{\bf The strong-CP question
in SU(3)$_c\times$SU(3)$_L\times$U(1)$_N$ models}}

\author{\bf Palash B. Pal\\
\normalsize \em Indian Institute of Astrophysics, Bangalore 560034,
             INDIA}

\date{}
\maketitle

\begin{abstract} \normalsize\noindent
We analyze two recent models based on the gauge
group SU(3)$_c\times$SU(3)$_L\times$U(1)$_N$ where each generation is
not anomaly-free, but anomaly cancels when three generations are taken
into account. We show that the most general Yukawa couplings of these
models admit of a Peccei-Quinn symmetry. This symmetry can be extended
to the entire Lagrangian by using extra fields in a very elegant way
so that the resulting axion can be made invisible.
\end{abstract}
\bigskip\bigskip

\subsection*{Introduction}
In the standard model, each generation of fermions is anomaly-free.
This is true for many extensions of the standard model as well,
including popular grand unified models. In these models, therefore,
the number of generations is completely unrestricted on theoretical
grounds. Recently, an interesting class of models have been proposed
which do not have this property. Each generation is anomalous, but
different generations are not exact replicas of one another, and the
anomalies cancel when a number of generations are taken into account.
The most economical gauge group which admits of such fermion
representations is SU(3)$_c\times$SU(3)$_L\times$U(1)$_N$, which has
been proposed by Pisano and Pleitez \cite{PiPl92} and by Frampton
\cite{Fra92} (PPF). The
original model did not have right-handed neutrinos, but
recently Foot, Long and Tran (FLT) \cite{FLT94} have included them in a
non-trivial way in an interesting variation of the model. Here we
show that the Yukawa couplings of these models automatically contains
a Peccei-Quinn (PQ) symmetry. The symmetry can also be extended to the
Higgs potential, thereby making it a symmetry of the entire
Lagrangian. This solves the strong CP problem in an elegant way in
these models. However, the resulting axion can be made consistent with
known bounds only by introducing new fields.

\subsection*{The PPF model}
The fermion representations of the Pisano-Pleitez-Frampton (PPF)
model are as follows~\cite{PiPl92,Fra92}:
	\begin{eqnarray}
f_{aL} & = \left( \begin{array}{c} e^+_a \\ \nu_a\\ e_a
\end{array} \right)_L & \sim (1,3,0) \\
Q_{1L} & = \left( \begin{array}{c} T_1 \\ u_1 \\ d_1
\end{array} \right)_L & \sim (3,3,{2\over3}) \\
Q_{iL} & =  \left( \begin{array}{c} u_i \\ d_i \\ B_i
\end{array} \right)_L & \sim (3,\bar 3,-{1\over3}) \\
u_{aR} && \sim (3,1,{2\over3}) \\
d_{aR} && \sim (3,1,-{1\over3}) \\
T_{1R} && \sim (3,1,{5\over3}) \\
B_{iR} && \sim (3,1,-{4\over3}) \,.
	\end{eqnarray}
The generation indices are of
two types: $a$ goes from 1 to 3, whereas $i$ takes only
the values 2 and 3.
Notice that in addition to the ordinary quarks, there are exotic ones,
with charges $5\over3$ and $-{4\over3}$. It is straightforward to
check that gauge anomalies cancel in this model.

In order to break the symmetry as well as to give masses to the
fermions, the following Higgs boson representations are needed:
	\begin{eqnarray}
\chi & = \left( \begin{array}{c} \chi_0 \\ \chi_- \\ \chi_{--}
\end{array} \right) & \sim (1,3,-1)\\
\rho & = \left( \begin{array}{c} \rho_{++} \\ \rho_+ \\ \rho_0
\end{array} \right) & \sim (1,3,1) \\
\eta & = \left( \begin{array}{c} \eta_+ \\ \eta_0 \\ \eta_-
\end{array} \right) & \sim (1,3,0) \\
S && \sim (1,6,0) \,.
	\end{eqnarray}
The most general Yukawa couplings consistent with gauge symmetry can
now be written as:
	\begin{eqnarray}
{\cal L}_Y &=& \phantom{+}
h_1 \overline Q_{1L} T_{1R} \chi + h_{2ij} \overline Q_{iL} B_{jR}
\chi^* \nonumber\\*
&& + h_{3a} \overline Q_{1L} d_{aR} \rho
+ h_{4ia} \overline Q_{iL} u_{aR} \rho^* \nonumber\\*
&& + h_{5a}\overline Q_{1L} u_{aR} \eta +
h_{6ia}\overline Q_{iL} d_{aR} \eta^* +
{\cal G}_{ab} f_{aL} f_{bL} \eta +
{\cal G}'_{ab} f_{aL} f_{bL} S^* \,.
\label{ppf.yuk}
	\end{eqnarray}
Notice that the introduction of the sextet $S$ is not essential for
the symmetry breaking. In fact, it is easy to see that the gauge symmetry
breaks to SU(3)$_c\times$U(1)$_Q$ if the triplet Higgs multiplets obtain
the following vaccum expectation values (VEVs):
	\begin{eqnarray}
\langle \chi_0 \rangle = v_\chi \,, \quad
\langle \rho_0 \rangle = v_\rho \,, \quad
\langle \eta_0 \rangle = v_\eta \,.
\label{ppf.vev}
	\end{eqnarray}
However, in this case, the mass matrix of the charged leptons would be
antisymmetric, and for three generations one eigenvalue will be zero
and the other two equal in magnitude. This is not realistic, and
therefore a VEV of the sextet is needed to produce a realistic mass
matrix~\cite{Fra92,FHPP93}.

We now show that the Yukawa couplings of Eq.\ (\ref{ppf.yuk}) respect
an extra global U(1) symmetry:
	\begin{eqnarray}
\begin{tabular}{l|ccccccc}
Multiplet & $\chi$ & $\eta$ & $\rho$ & $S$ & $Q_{1L}$ & $Q_{iL}$
& $f_{aL}$ \\  \hline
U(1) charge & 1 & 1 & 1 & 2 & 1 & $-1$ & 1\\
\end{tabular}
\label{ppf.PQ}
	\end{eqnarray}
Since the charges of left and right chiral fermions are unequal, this
is a chiral symmetry, of the type envisaged by Peccei and Quinn
\cite{PQ77} to solve the strong CP problem. If this symmetry can be
extended to the entire Lagrangian, the model can be made free from the
strong CP problem.

The most general Higgs potential with these Higgs multiplets is given
by~\cite{FHPP93}:
	\begin{eqnarray}
V(\chi,\rho,\eta,S) &=& \phantom{+}
\lambda_1 (\chi^\dagger\chi - v_\chi^2)^2 +
\lambda_2 (\rho^\dagger\rho - v_\rho^2)^2 +
\lambda_3 (\eta^\dagger\eta - v_\eta^2)^2 \nonumber\\* &&
+ \lambda_4 [\mbox{tr}\; (S^\dagger S) - v_S^2]^2
+ \lambda_5 [\mbox{tr}\; (S^\dagger SS^\dagger S) - v_S^4]
\nonumber\\*  &&
+ \lambda_6 (\chi^\dagger\chi \rho^\dagger\rho - v_\chi^2 v_\rho^2)
+ \lambda_7 (\rho^\dagger\rho \eta^\dagger\eta - v_\rho^2 v_\eta^2)
+ \lambda_8 (\eta^\dagger\eta \chi^\dagger\chi - v_\eta^2 v_\chi^2)
\nonumber\\* &&
+ \lambda_9 (\chi^\dagger\chi \;\mbox{tr}\, (S^\dagger S)
- v_\chi^2 v_S^2)
+ \lambda_{10} (\rho^\dagger\rho \;\mbox{tr}\, (S^\dagger S)
- v_\rho^2 v_S^2)
+ \lambda_{11} (\eta^\dagger\eta \;\mbox{tr}\, (S^\dagger S)
- v_\eta^2 v_S^2) \nonumber\\* &&
+ \lambda_{12} \chi^\dagger\rho \rho^\dagger\chi
+ \lambda_{13} \rho^\dagger\eta \eta^\dagger\rho
+ \lambda_{14} \eta^\dagger\chi \chi^\dagger\eta \nonumber\\* &&
+ \mu_1 \eta\rho\chi + \mu_2 \rho^T S^\dagger \chi + \mbox{h.c.}
	\end{eqnarray}
If the PQ symmetry is imposed on the
Higgs potential as well, the trilinear interaction
term $\eta\rho\chi$ drops out. However, from the
analysis of the Higgs potential \cite{FHPP93}, it is easy to see
that one still obtains the vacuum given in Eq.\ (\ref{ppf.vev}).

In fact, it is not even necessary to impose the PQ symmetry
to get rid of the trilinear term mentioned above. One can, for
example, use a discrete symmetry
	\begin{eqnarray}
\chi \to -\chi \,, \quad T_{1R} \to - T_{1R} \,, \quad B_{iR} \to -
B_{iR} \,.
\label{ppf.dsymm}
	\end{eqnarray}
Then the aforementioned term is automatically eliminated, and
PQ symmetry emerges as an automatic symmetry of the
classical Lagrangian.

This approach has an advantage. Usually the
imposition of the PQ symmetry is somewhat awkward. One has
to impose it at the classical level only, but the quantum corrections
break it through instanton effects.
There has been a number of attempts to find models where
the PQ symmetry follows as a natural consequence of other
symmetries at the classical level \cite{autoPQ}. This
model seems to be one of that sort if the discrete symmetry of Eq.\
(\ref{ppf.dsymm}) is imposed.

However, this also shows that the model, as described above, cannot be
realistic. This is because the process of symmetry breaking breaks the
PQ symmetry spontaneously, and an axion results. Naively, it seems
from Eq.\ (\ref{ppf.PQ}) that the spontaneous breaking of the PQ
symmetry takes place at the scale $v_\chi$, which can be much higher
than the weak scale, and therefore the axion can be invisible. This is
not true. At the scale $v_\chi$, although the U(1)$_{\rm PQ}$ breaks,
the combination $\rm U(1)_{PQ}+U(1)_N$ remains unbroken, and this acts
as the PQ symmetry. This symmetry is broken at the weak scale by the
VEVs of $\eta$ and $\rho$, so that the axion is of
the Weinberg-Wilczek type \cite{WWaxion},
which has been ruled out by experiments.
One therefore, must extend this model in order to make the axion
invisible.

The standard way of achieving this goal \cite{invaxion} is
to introduce a gauge
singlet field $\Phi$ which has a non-zero charge under the U(1)$_{\rm
PQ}$ symmetry. This field can then develop a large VEV to break the PQ
symmetry at a very high scale, thus making the axion invisible.
The problem is that, from gauge symmetry
requirements alone, in
addition to terms like $|\Phi|^2$, $|\Phi|^4$ which are
invariant under a U(1) phase of the $\Phi$-field,
the Higgs potential can also have
terms like $\Phi^2$, $\Phi^3$, $\Phi^4$ etc. In this case, one
has to impose the PQ symmetry on the classical Lagrangian.
As we argued above, this is a somewhat awkward procedure.

In this case, however, other possibilities remain. For example,
instead of introducing an extra gauge singlet, we can introduce a
Higgs boson multiplet
	\begin{eqnarray}
\Delta \sim (1,10,-3) \,,
	\end{eqnarray}
This multiplet does not
have any Yukawa coupling. In the Higgs potential, apart from the
pure-$\Delta$ terms and the terms which involve the combination
$\Delta^\dagger\Delta$, only the following term can be present in the
Higgs potential:
	\begin{eqnarray}
\lambda_{15} \chi^3\Delta^* + \mbox{h.c.} \,,
	\end{eqnarray}
provided $\Delta$ is odd under the discrete symmetry of
Eq.\ (\ref{ppf.dsymm}). This term governs that the PQ charge
of the multiplet $\Delta$ should
be 3. Thus, the PQ symmetry need not be imposed on the classical
Lagrangian in this case, it follows automatically from gauge symmetry.
This $\Delta$ has one component which is a singlet of
SU(2)$_L\times$U(1)$_Y$. If this component develops a VEV, the PQ
symmetry will be broken. If this symmetry is broken at a high scale,
above $10^9$~GeV, the resulting axion will be invisible.

\subsection*{The FLT model}
Foot, Long and Tran modified the above model to include right-handed
neutrinos. In their model \cite{FLT94}, the fermions
appear in the following gauge multiplets:
	\begin{eqnarray}
f_{aL} & = \left( \begin{array}{c} \nu_a \\ e_a \\ \widehat \nu_a
\end{array} \right)_L & \sim (1,3,-{1\over3}) \\
e_{aR} && \sim (1,1,-1) \\
Q_{1L} & = \left( \begin{array}{c} u_1 \\ d_1 \\ u'_1
\end{array} \right)_L & \sim (3,3,{1\over3}) \\
Q_{iL} & =  \left( \begin{array}{c} d_i \\ u_i \\ d'_i
\end{array} \right)_L & \sim (3,\bar 3,0) \\
u_{aR}, u'_{1R} && \sim (3,1,{2\over3}) \\
d_{aR}, d'_{iR} && \sim (3,1,-{1\over3}) \,.
	\end{eqnarray}
Here, $\widehat\nu$ denotes antineutrinos, and $u'$ and $d'$ are new
quark fields introduced in the model.
 Thus, like the previous model,
one generation of left-handed quarks transform differently from
the other ones. The representations of the leptons are different from
the earlier model, which includes the right-handed neutrinos in the
triplet. Choice of this representation also allows FLT
\cite{FLT94} to eliminate quark fields of exotic charges, and
to use a more economical Higgs boson sector which
performs the necessary gauge symmetry breaking:
	\begin{eqnarray}
\chi & = \left( \begin{array}{c} \chi_0 \\ \chi_- \\ \chi'_0
\end{array} \right) & \sim (1,3,-{1\over3}) \\
\rho & = \left( \begin{array}{c} \rho_+ \\ \rho_0 \\ \rho'_+
\end{array} \right) & \sim (1,3,{2\over3}) \\
\eta & = \left( \begin{array}{c} \eta_0 \\ \eta_- \\ \eta'_0
\end{array} \right) & \sim (1,3,-{1\over3}) \,.
	\end{eqnarray}
It should be noted that the gauge group and the representations
mentioned here are identical to an earlier model by Singer, Valle and
Schechter (SVS) \cite{SVS80}. The only difference
is that the third member
of the lepton triplet was interpreted as a new neutrino field by SVS,
and they had to introduce its right handed partner as well, which is a
gauge singlet. Here, since the third member is interpreted as the
antineutrino, no extra leptonic fields are necessary.

The most general Yukawa couplings of the model can be
written as:
	\begin{eqnarray}
{\cal L}_Y &=& \phantom{+} h_1 \overline Q_{1L} u'_{1R} \chi +
h_{2ij} \overline Q_{iL} d'_{jR} \chi^* \nonumber\\*
&&+ h_{3a} \overline Q_{1L} u_{aR} \eta +
h_{4ia} \overline Q_{iL} d_{aR} \eta^* \nonumber\\*
&&+ h_{5a} \overline Q_{1L} d_{aR} \rho +
h_{6ia} \overline Q_{iL} u_{aR} \rho^* +
{\cal G}_{ab} f_{aL} f_{bL} \rho +
{\cal G}'_{ab} \overline f_{aL} e_{bR} \rho + \mbox{h.c.} \,.
	\end{eqnarray}
FLT \cite{FLT94} showed that if the
vacuum expectation values (VEVs) of the neutral scalar fields are
given by:
	\begin{eqnarray}
\langle \chi'_0 \rangle = v_\chi \,, \quad
\langle \rho_0 \rangle = v_\rho \,, \quad \label{vevs}
\langle \eta_0 \rangle = v_\eta \,,
	\end{eqnarray}
where all others have zero VEV, the symmetry breaks to
SU(3)$_c\times$U(1)$_Q$, and the fermions obtain masses.

We now show that the Yukawa couplings of this model respect an extra
global U(1) symmetry. The charges of various multiplets under this
symmetry are as follows:
	\begin{eqnarray}
\begin{tabular}{l|ccccccc}
Multiplet & $\chi$ & $\eta$ & $\rho$ & $Q_{1L}$ & $Q_{iL}$
& $f_{aL}$ & $e_{aR}$  \\ \hline
U(1) charge & 1 & 1 & 1 & 1 & $-1$ & $-{1\over2}$ & $-{3\over 2}$\\
\end{tabular}
\label{PQ}
	\end{eqnarray}
all other multiplets being neutral. Once again, it is obvious that
this is a chiral Peccei-Quinn \cite{PQ77} symmetry.
The present model, then, has the property that it does not
suffer from the strong CP problem if this symmetry can be extended to
the entire Lagrangian.

The Higgs potential advocated by
FLT \cite{FLT94} is not the most general one subject to gauge
invariance. In fact, they imposed a discrete
symmetry on the Lagrangian:
	\begin{eqnarray}
\chi \to -\chi \,, \qquad u'_{1R} \to - u'_{1R} \,, \qquad
d'_{jR} \to - d'_{jR} \,.
\label{dsymm}
	\end{eqnarray}
All the terms in the Yukawa couplings given above are allowed under
this symmetry. The Higgs potential allowed under this symmetry
is~\cite{FLT94}:
	\begin{eqnarray}
V(\chi,\eta,\rho) &=& \phantom{+}
\lambda_1 (\chi^\dagger \chi - v_\chi^2)^2 +
\lambda_2 (\eta^\dagger \eta - v_\eta^2)^2 +
\lambda_3 (\rho^\dagger \rho - v_\rho^2)^2 \nonumber\\*
&& +
\lambda_4 (\chi^\dagger \chi - v_\chi^2) (\eta^\dagger \eta - v_\eta^2) +
\lambda_5 (\eta^\dagger \eta - v_\eta^2) (\rho^\dagger \rho - v_\rho^2) +
\lambda_6 (\rho^\dagger \rho - v_\rho^2) (\chi^\dagger \chi - v_\chi^2)
\nonumber\\*
&&
+ \lambda_7 (\chi^\dagger\eta + \eta^\dagger\chi)^2
+ \lambda_8 (\chi^\dagger\eta) (\eta^\dagger\chi)
+ \lambda_9 (\eta^\dagger\rho) (\rho^\dagger\eta)
+ \lambda_{10} (\rho^\dagger\chi) (\chi^\dagger\rho) \,.
	\end{eqnarray}
Now, if $\lambda_1,\cdots,\lambda_{10}\geq 0$, the
vacuum structure mentioned in Eq.\ (\ref{vevs}) can be obtained
from the minimization of this potential \cite{FLT94}.
The first six terms imply that the magnitude of the VEVs are
$v_\chi$, $v_\eta$ and $v_\rho$, whereas the last four terms imply
that they must be in orthogonal directions.

It is now trivial to see that the U(1)$_{\rm PQ}$ symmetry given in
Eq.\ (\ref{PQ}) is a symmetry of the Higgs potential as well. Thus, it
is a symmetry of the entire Lagrangian. This is an automatic symmetry,
in the sense that it does not have to be imposed separately on the
Lagrangian. Rather, it comes as a consequence of the gauge symmetry
and the discrete symmetry of Eq.\ (\ref{dsymm}).

Unfortunately, like the previous model, here also the symmetry is
broken by the instanton effects as well as spontaneously, and
therefore an axion results.
This spontaneous breaking of the U(1)$_{\rm PQ}$ occurs at
the weak scale by the VEVs of $\eta$ and $\rho$. Therefore, the axion
is of the Weinberg-Wilczek type \cite{WWaxion}, and therefore again
the model needs to be extended in order to make it realistic.

Here also, one can get rid of the problem by
introducing a gauge singlet Higgs boson \cite{invaxion} which
transforms non-trivially under the PQ symmetry. But in this
case, the PQ symmetry does not arise automatically, and we argued the
awkwardness of this procedure.

Fortunately, in this case as well, one can do otherwise. One can
introduce a Higgs boson multiplet
	\begin{eqnarray}
\Delta \sim (1,10,-1) \,.
	\end{eqnarray}
The non-trivial couplings of this field are
	\begin{eqnarray}
\lambda_{11} \eta^3 \Delta^* + \lambda_{12} \eta\chi^2 \Delta^*
	\end{eqnarray}
provided $\Delta$ is even under the discrete symmetry of Eq.\
(\ref{dsymm}), and
	\begin{eqnarray}
\lambda'_{11} \chi^3 \Delta^* + \lambda'_{12} \eta^2\chi \Delta^*
	\end{eqnarray}
provided $\Delta$ is odd. In either case, these couplings dictate that
the PQ charge of the multiplet $\Delta$ should be 3. When the standard
model singlet component of $\Delta$ develops a VEV alongwith the VEV
of $\chi$, the PQ symmetry is broken. If this symmetry breaking takes
place above $10^9$ GeV, the axion is invisible.

\subsection*{Conclusions}
We have thus shown the in recently proposed models based on the gauge
group SU(3)$_c\times$SU(3)$_L\times$U(1)$_N$, the strong CP problem
can be solved in a very elegant way. The elegance comes from the fact
that the PQ symmetry does not have to be {\em imposed} in these models.
Rather, with a judicious choice of the Higgs boson multiplets, the PQ
symmetry can follow automatically from the gauge invariant Lagrangian.
We explained that this is a very satisfying feature, since the
imposition of the PQ symmetry is awkward at the classical level,
knowing that it is anyway broken by instanton effects when quantum
effects are considered.

However, in order to achieve this, we needed to extend the Higgs boson
sector of the model, which can be considered as an ugly feature. More
importantly, our implementation of the PQ symmetry with the decouplet
fields imply that the SU(3)$_L$ symmetry has to be broken at a rather
high scale, higher than $10^9$ GeV. Thus, the massive gauge bosons
which occur in this model, apart from the $W^\pm$ and the $Z$, are all
very heavy, and the model loses a lot of its rich phenomenological
consequences which would have occurred if these gauge bosons were much
lighter. If the Peccei-Quinn symmetry is broken by a gauge singlet
Higgs boson field, then of course the gauge bosons can be much
lighter, because only the global PQ symmetry is broken at high scale
and not gauge symmetry, but in this case one has to live with the
awkward implementation of the PQ symmetry.

\end{document}